\documentclass[a4paper,fleqn,usenatbib]{mnras}
\usepackage[T1]{fontenc}
\usepackage{ae,aecompl}
\usepackage{graphicx}	
\usepackage{amsmath}	
\usepackage{amssymb}	

\title[radial distribution of stellar populations in NGC$\,$2808]{The {\it Hubble Space Telescope} UV Legacy Survey of Galactic Globular Clusters. X. The radial distribution of stellar populations in NGC$\,$2808 \thanks{Based on observations with the NASA/ESA {\it Hubble Space Telescope}, obtained at the Space Telescope Science Institute, which is operated by AURA, Inc., under NASA contract NAS 5-26555.}}

\author[M.\,Simioni et al.]{
M.\,Simioni$^{1,2,3,4}$\thanks{email: msimioni@iac.es},
A.\,P.\,Milone$^{5}$,
L.\,R.\,Bedin$^{4}$,
A.\,Aparicio$^{2,1}$,
G.\,Piotto$^{3,4}$,\newauthor
E.\,Vesperini$^{6}$,
J.\,Hong$^{6}$
\\
$^{1}$Instituto de Astrof\`isica de Canarias, E-38200 La Laguna, Tenerife, Canary Islands, Spain\\
$^{2}$Department of Astrophysics, University of La Laguna, E-38200 La Laguna, Tenerife, Canary Islands, Spain\\
$^{3}$Dipartimento di Fisica e Astronomia ``Galileo Galilei'', Universit\`{a} degli Studi di Padova,  Vicolo dell'Osservatorio 3, Padova IT-35122\\
$^{4}$INAF - Osservatorio Astronomico di Padova, Vicolo dell'Osservatorio 5, I-35122 Padova, Italy\\
$^{5}$Research School of Astronomy and Astrophysics, The Australian National University, Cotter Road, Weston, ACT, 2611, Australia\\
$^{6}$Department of Astronomy, Indiana University, Bloomington, IN47401, USA\\
}

\date{Accepted XXX. Received YYY; in original form ZZZ}

\pubyear{2016}

\begin{document}
\label{firstpage}
\pagerange{\pageref{firstpage}--\pageref{lastpage}}
\maketitle

\begin{abstract}  
 Due to their extreme helium abundance, the multiple stellar populations of the globular cluster NGC\,2808 have been widely investigated from a photometric, spectroscopic, and kinematic perspective. 
 The most striking feature of the color-magnitude diagram of NGC\,2808 is the triple main sequence (MS), with the red MS corresponding to a stellar population with primordial helium, and the middle and the blue MS being enhanced in helium up to $Y \sim$0.32 and $\sim$0.38, respectively. 
 A recent study has revealed that this massive cluster hosts at least five distinct stellar populations (A, B, C, D, and E). Among them populations A, B, and C correspond to the red MS, while populations C and D are connected to the middle and the blue MS.  
 In this paper we exploit {\it Hubble-Space-Telescope} photometry to investigate the radial distribution of the red, the middle and the blue MS from the cluster center out to about 8.5 arcmin.
 Our analysis shows that the radial distribution of each of the three MSs is different. In particular, as predicted from multiple-population formation models, both the blue MS and the middle MS appears to be more concentrated than the red MS with a significance level for this result wich is above $3\sigma$.
\end{abstract}


\defcitealias{2012A&A...537A..77M}{M12}
\defcitealias{2015ApJ...808...51M}{Paper\,III}
\defcitealias{2015AJ....149...91P}{Paper\,I}
\section{Introduction}\label{sec:intro}
The massive globular cluster (GC) NGC\,2808 is one of the most intriguing objects in the context of multiple stellar populations.
The most astonishing feature of its color-magnitude diagram (CMD) is the presence of five distinct sequences of main-sequence (MS), and red-giant-branch (RGB) stars (\citealt{2007ApJ...661L..53P}; \citealt{2015ApJ...808...51M} -- hereafter \citetalias{2015ApJ...808...51M} --, \citealt{2012A&A...537A..77M} --hereafter \citetalias{2012A&A...537A..77M}--) and at least four distinct horizontal-branch (HB) segments \citep{2000A&A...363..159B}.

 Spectroscopy of bright RGB stars has revealed an extreme chemical composition  with extended Na-O (\citealt{2006A&A...450..523C}; \citealt{2013A&A...549A..41G}; \citealt{2014MNRAS.437.1609M}) and Mg-Al \citep{2014ApJ...795L..28C} anticorrelations.

  The distinct sequences in the CMD of NGC\,2808 correspond to multiple stellar populations with light element abundance variations and different helium content. 
 In particular, the three most evident MSs discovered by \citet{2007ApJ...661L..53P}, namely red, middle, and blue MS (rMS, mMS, and bMS) have been interpreted with three stellar populations with primordial helium abundance (Y$\sim$0.25) and with extreme values of Y$\sim$0.32, Y$\sim$0.38 (\citealt{2004ApJ...611..871D,2005ApJ...631..868D}; \citealt{2007ApJ...661L..53P}; \citetalias{2015ApJ...808...51M}). Large helium enhancement have been also inferred from spectroscopy of HB stars (\citealt{2014MNRAS.437.1609M}) and by the analysis of chromospheric lines in spectra of RGB stars \citep{2011A&A...531A..35P}.

The formation and evolution of stellar populations in GCs have been widely investigated by several authors (see e.g. \citealt{2015MNRAS.454.4197R} and references therein). The fraction of stars in each population, their radial distribution, chemical composition, mass function and dynamics are amongst the diagnostics commonly used to constraint the various scenarios. 
In particular, the radial distribution of stellar populations can provide information on the series of events that led from massive clouds in the early Universe to the present day GCs with their multiple stellar populations. Indeed clusters with long relaxation times may still keep information of the initial conditions of their stellar populations. 

Theoretical models and simulations by \citet{2008MNRAS.391..825D,2010MNRAS.407..854D} predict that stars enhanced in helium and sodium are more centrally concentrated than stellar populations with primordial helium and oxygen abundance. This scenario is in agreement with observational studies on some GCs (e.g.\,47Tuc and $\omega$Cen, \citealt{2007ApJ...654..915S}, \citealt{2009A&A...507.1393B}, \citealt{2012ApJ...744...58M}, \citealt{2014ApJ...780...94C}; M2, M3, M5, M13, M15, M53, M92, \citealt{2011A&A...525A.114L}; NGC\,362, \citealt{2013A&A...557A.138C}; NGC\,3201, \citealt{2010A&A...519A..71C}; NGC\,2419, \citealt{2013MNRAS.431.1995B}; NGC6388 and NGC6441, \citealt{2013ApJ...765...32B}) while in other cases the multiple stellar populations share the same radial distribution (e.g.\,NGC\,1851, NGC\,6121 (M4), NGC\,6362, and NGC\,6752, see \citealt{2009A&A...503..755M}; \citealt{2014ApJ...791L...4D}; \citealt{2015A&A...573A..70N}). 

In this paper we exploit proprietary data from the Wide Field Channel of the Advanced Camera for Surveys (WFC/ACS) and the Ultraviolet and Visual Channel of the Wide Field Camera III (UVIS/WFC3) to investigate for the first time the radial distribution of the multiple MSs in NGC\,2808.
We also show the results of a simple N-body simulations aimed at illustrating and providing some insight on the possible spatial mixing and dynamical history of this cluster.
This paper is part of the {\it Hubble Space Telescope} ({\it HST\/}) UV Legacy Survey of Galactic Globular Clusters that is a project to investigate 57 Galactic Globular Clusters (GCs) through the filters F275W, F336W and F438W of UVIS/WFC3 (GO-13297, PI.\,G.\,Piotto, see \citealt{2015AJ....149...91P} -- hereafter \citetalias{2015AJ....149...91P} -- for details).
 The paper is organized as follows:
In Sect.~\ref{sec:data} we present the data and the data analysis; Sect.~\ref{sec:ms} describes in detail the methods used to derive the fraction of bMS, mMS, and rMS stars and the simulation performed for the theoretical analysis.
 Results are then presented in Sect.~\ref{sec:results}, and discussed in Sect.~\ref{sec:summary}.

\section{Data and data analysis}\label{sec:data}
 In order to study the radial distribution of the multiple MSs and RGBs of NGC\,2808 we have exploited the dataset listed in Tab.~\ref{tab:logobs} which consists of images taken with ACS/WFC and UVIS/WFC3 on board of {\it HST}. These data are part of GO-9899, GO-10922, GO-12605 (PI.\,G.\,Piotto) and GO-10775 (PI.\,A.\,Sarajedini) and most of them have been already used by our group to study multiple stellar populations in this clusters (e.g.\,\citealt{2007ApJ...661L..53P}, \citetalias{2015AJ....149...91P}; \citealt{2007AJ....133.1658S}; \citealt{2008AJ....135.2055A}; \citetalias{2012A&A...537A..77M}). 

\begin{figure}
\centering
   \includegraphics[width=9cm]{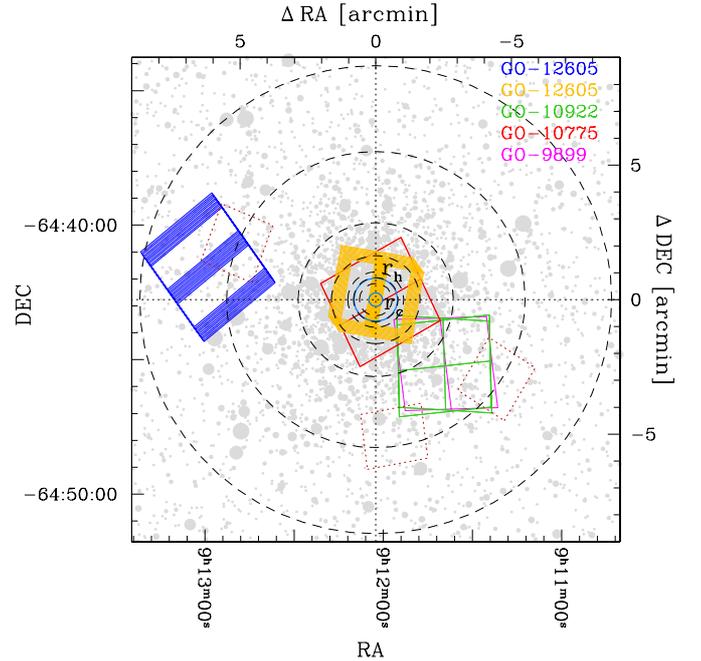}
   \caption{ Finding chart for the used images.  
Core and half-light radii of NGC\,2808 ($r_{c}=0.25$ arcmin and $r_{h}=0.80$ arcmin, \citealt{1996yCat.7195....0H}) are indicated by cyan circles. For each used set the corresponding footprints are shown: red dotted fields corresponds to WFC3 IR observations of GO-11665. Dashed circles represents the boudaries of the selected radial bins; see text for details.}
   \label{fig:chart}
\end{figure}

 The footprints of the data are shown in Fig.~\ref{fig:chart} where we indicate with different color codes the images from different GOs. Stars in the most external field have radial distance of $\Delta$r$\sim$8.5 arcmin at most and lie approximately halfway from the tidal radius of NGC\,2808 which is $r_{t}=9.08$ arcmin \citep{1996yCat.7195....0H}. 

We have used the photometric and astrometric catalogs presented by \citetalias{2012A&A...537A..77M} from the GO-9899 and GO-10922 dataset, and the catalog from \citet{2008AJ....135.2055A} for GO-10775. The photometric and astrometric reduction for GO-12605 data has been carried out as described below.

We have first corrected the charge-transfer efficiency (CTE) effects in each image by using the method and the software by \citet{2010PASP..122.1035A}.

Photometry and astrometry of ACS/WFC images has been  performed as in \citet{2008AJ....135.2055A}.
Briefly, we have used two distinct methods to measure bright and faint stars. To determine flux and position of bright stars we have analyzed each image, independently, by using the point-spread function models from \citet{2006acs..rept....1A} plus a spatially constant perturbation that accounts for small focus variations due to the `breathing' of {\it HST}. The derived magnitudes and positions are then combined. The flux and the position for very faint stars, which can not be robustly measured in every individual image, have been determined by fitting for each star simultaneously all the pixels in all the exposures (see Section 5 of \citealt{2008AJ....135.2114A} for more details).  Stellar positions have been corrected for geometrical distortion by using the solution provided by \citet{2006acs..rept....1A}.
UVIS/WFC3 images have been analysed similarly. In this case, we derived the PSFs as in \citet{2006A&A...454.1029A} and \citet{2010AJ....140..631B} and used distortion solution by \citet{2009PASP..121.1419B} and \citet{2011PASP..123..622B}.

We have used the several indexes provided by the software as diagnostics of the quality of photometry \citep{2008AJ....135.2055A}. Since high-accuracy photometry is required to analyse the multiple MSs and RGBs in NGC\,2808, we have adopted the method described by \citet{2009A&A...497..755M} and \citet{2009ApJ...697..965B} to select a sub-sample of stars that have small astrometric errors, are relatively isolated, and well fitted by the PSF as in \citet{2009A&A...497..755M} (Sect.\,2.1).

Photometry has been calibrated as in \citet{2005MNRAS.357.1038B}. For the WFC/ACS images, we used the zero points provided by Bedin and collaborators, while for UVIS/WFC3 we adopted the zero points listed in the STScI web page for WFC/ACS and WFC3/UVIS\footnote{http://www.stsci.edu/hst/wfc3/phot\_zp\_lbn, http://www.stsci.edu/hst/acs/analysis/zeropoints/zpt.py}.  Each CMD has been corrected for differential reddening as in \citetalias{2012A&A...537A..77M}.

\begin{figure*}
\centering
   \includegraphics[width=12cm]{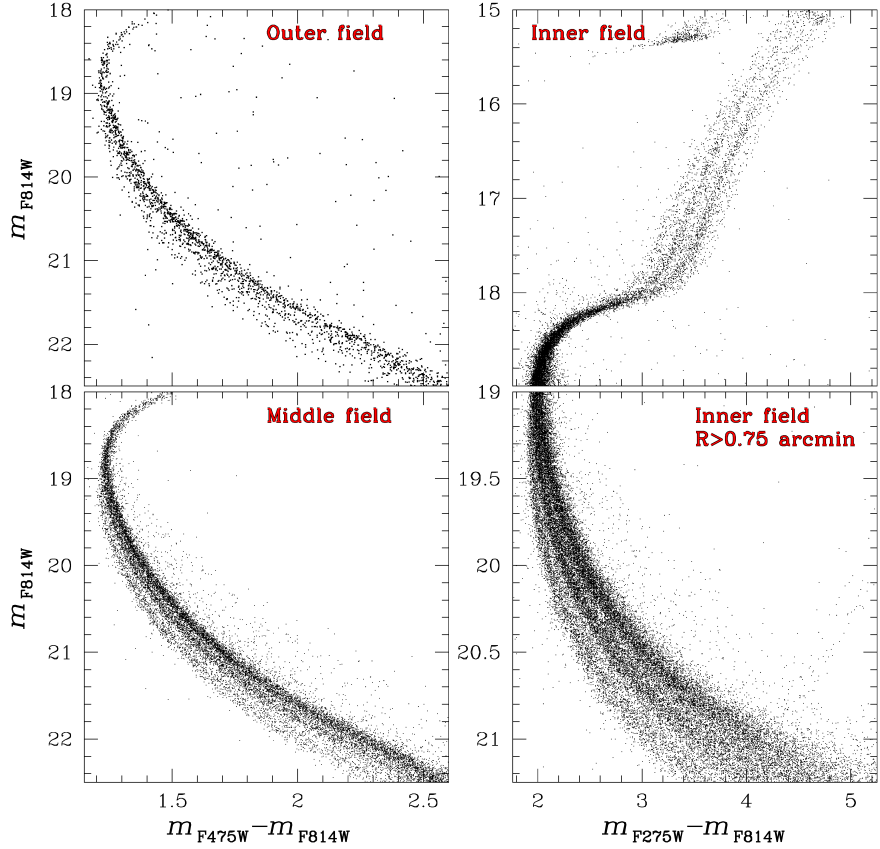}
   \caption{CMDs corrected for differential reddening used to determine the fraction of stars in the multiple MSs of  NGC\,2808 in the outer and middle field (left panels), and the fraction of multiple MSs and RGBs in the inner field (right panels).}
   \label{fig:CMDs}
\end{figure*}

The CMDs used to study the radial gradient of multiple populations in NGC\,2808 are shown in Fig.~\ref{fig:CMDs}. For the central and middle field where stellar proper motions were available from \citet{2007ApJ...661L..53P} and \citetalias{2015ApJ...808...51M} we analyzed only stars that, according to their motion are cluster members (see \citealt{2007ApJ...661L..53P} for details). On the contrary, there are no proper motions for stars in the outer field and we will account for field-star contamination by using the Galactic model from \citet{2005A&A...436..895G} as discussed in Sect.~\ref{sec:ms}.  
 In the outer and middle field we have analyzed the $m_{\rm F814W}$ vs.\,$m_{\rm F475W}-m_{\rm F814W}$ CMD and studied multiple populations in the magnitude interval $19.50<m_{\rm F814W}<22.00$ where the three MSs are clearly visible. In the inner field we analyzed MS stars with $19.50<m_{\rm F814W}<21.25$ and limited our study to the region with distance from the cluster center, $R>$0.75 arcmin, indeed crowding prevents us to clearly distinguish the triple MS at smaller radii. In this case we used the $m_{\rm F814W}$ vs.\,$m_{\rm F275W}-m_{\rm F814W}$ CMD. Furthermore, we analyze the radial distribution of multiple populations along the RGB in the entire inner field. 

\begin{table*}\footnotesize
  \centering
\scriptsize {
    \begin{tabular}{@{}ccccccc@{}} 
    \hline
    GO & PI & Camera & Filter & Exposures & $R$ & Epoch \\
    \hline
    9899  & G.\,Piotto &  WFC/ACS & F475W & 6$\times$340s & $3'.34$ & 05 May 2004 \\ 
    10775 & A.\,Sarajedini & WFC/ACS & F814W & 23s$+$5$\times$370s & & 01 Jan 2006 \\ 
    10922 & G.\,Piotto & WFC/ACS & F475W & 2$\times$350s & $3'.40$ & 09 Aug 2006\\
    {}    & {}     &    & {} & 2$\times$360s & $3'.38$ & 01 Nov 2006\\
    {}    & {}     &    & F814W & 3$\times$350s & $3'.40$ & 09 Aug 2006\\
    {}    & {}     &    & {} & 3$\times$360s & $3'.38$ & 01 Nov 2006\\ 
    12605 & G.\,Piotto & WFC/ACS & F475W & 6$\times$890s$+$6$\times$982 s & $6'.27$ & 08 Sept 2013\\
    {} &  & {} & F814W & 6$\times$508s & $6'.27$ & 08-9 Sp 2013\\
    12605 & G.\,Piotto & UVIS/WFC3 & F275W & 12$\times$985s & & 08-09 Esp 2013  \\
    \hline
   \end{tabular}
    }
   \label{tab:logobs}
 \caption{ List of the used {\it HST} images. The average radial distance from the clusters center, $R$, is indicated for each field. }
\end{table*}

\subsection{Artificial Stars} 
 The artificial-star (AS) experiments have been performed as in \citet{2008AJ....135.2055A}.
We have generated a list of 300,000 stars and placed them along each MS and RGB of NGC\,2808 by assuming the same spatial distribution as observed for real stars (see \citetalias{2012A&A...537A..77M} for details). 

For each star in the input list, we have generated, a star in each image, and measures it by using the same procedure as for real stars. We assumed an AS as found when the input and the output position differ by less then 0.5 pixel and the input and the output flux by less than 0.75 mag. 
Finally, we have selected a sub-sample of relatively isolated ASs with small astrometric errors, and well fitted by the PSF by using the same procedure as for real stars indeed the software by \citet{2008AJ....135.2055A} provides for ASs the same diagnostics of the photometric quality as for real star.

ASs have been used to estimate errors of the photometry used in this paper and the completeness level.
Completeness have been derived for each star as in \citetalias{2012A&A...537A..77M}.

\section{The fraction of stars in the three MSs}
\label{sec:ms}

\begin{centering}
\begin{figure*}
 \includegraphics[width=12.5cm]{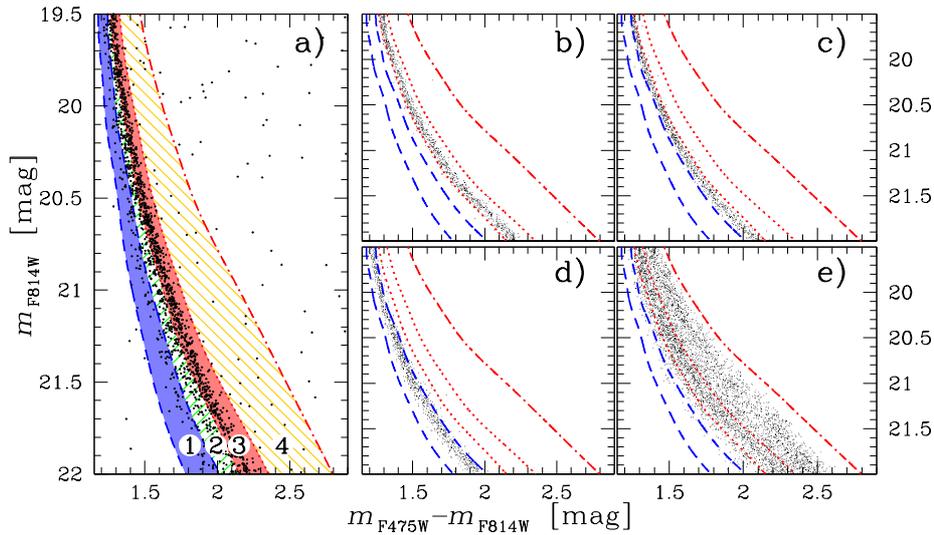}
 \caption{Method used to determine the fraction of bMS, mMS and rMS stars and the fraction of binaries from GO-12625 ACS/WFC data. Blue-dashed and red-dotted lines have been obtained by shifting the fiducials of the bMS and the rMS by $\pm\,3\,\sigma_{\rm bMS}$ and $\pm\,3\,\sigma_{\rm rMS}$, respectively. The red dashed-dotted line plotted in each panel is the fiducial line of equal-mass binaries red-shifted by $3\,\sigma_{\rm rMS}$. These lines are the boundaries of four regions, R$_{1}$--R$_{4}$ of the CMD colored blue, green, red, and yellow, respectively in panel a). The observed CMD for NGC\,2808 stars is plotted in panel a), while panels b), c), and d) show the simulated CMD for single rMS, mMS, and bMS stars, respectively. The CMD of panel e) is made only of MS-MS binaries (see text for details).}
 \label{fig:setup}
\end{figure*}
\end{centering}

To compare the radial distribution of each stellar population and highlight the presence of gradients among them, the number of stars in each MS has been counted. These values have been thus normalized to the total number of MS stars observed in the analyzed magnitude ranges in order to obtain the so-called population ratio.
To derive the number of stars in each MS ($N_{\rm bMS}$, $N_{\rm mMS}$, and $N_{\rm rMS}$) and the fraction of binaries ($f^{\rm BIN}$) in each field of NGC\,2808 we adopted the iterative procedure introduced by \citetalias{2012A&A...537A..77M} and illustrated in Fig.~\ref{fig:setup} for the CMD of the outer field. The blue dashed lines and the red dotted lines plotted in each panel of Fig.~\ref{fig:setup} are obtained by shifting by plus or minus $3\,\sigma_{\rm bMS}$ and $3\,\sigma_{\rm rMS}$ the fiducial line of the red and the blue MS, respectively. The red dashed-dotted line is the fiducial line of equal mass rMS-rMS binaries red-shifted by $3\,\sigma_{\rm rMS}$. The fiducials and the corresponding errors have been determined as in \citetalias{2012A&A...537A..77M}.

These lines are the boundaries of four regions that we name $R_{1}$, $R_{2}$, $R_{3}$, and $R_{4}$ and colored blue, green, red, and yellow, respectively.
 The majority of the bMS, mMS, and rMS stars are located in the regions $R_{1}$, $R_{2}$, and $R_{3}$, while the region $R_{4}$ is mostly  populated by MS-MS binaries. As discussed by \citetalias{2012A&A...537A..77M}, a fraction of bMS also migrates into regions $R_{2, 3, 4}$, and similar shift applies to some mMS and rMS stars. Moreover regions $R_{1}$, $R_{2}$, $R_{3}$ are also populated by MS-MS binaries.   
 
Specifically, each region $R_{\rm i}$ is populated by a fraction $f_{\rm i}^{\rm bMS (mMS, rMS)}$ of bMS (mMS, rMS) stars and a fraction $f_{\rm i}^{\rm BIN}$ of binaries. The relations between the observed total numbers of stars  and  the number of stars in each sequence  can be expressed, for i=1,2,3,4 as:

\begin{equation}
\label{eq:1}
  N_{\rm i} = N_{\rm bMS} f^{\rm bMS}_{\rm i} + N_{\rm mMS} f^{\rm
    mMS}_{\rm i} + N_{\rm rMS} f^{\rm rMS}_{\rm i} + f^{\rm BIN}
    N_{\rm MS} f^{\rm BIN}_{\rm i}
\end{equation}
 
where $N_{\rm MS}$ is the total number of MS stars.
$N_{\rm i}$ is the number of cluster stars in each region corrected for completeness.
 In the inner and middle field, where stellar proper motions are available, we have determined $N_{\rm i}$ from counts of stars that, according to their proper motions, are cluster members.
 In the outer field, where we do not have proper motions, we have used the Galactic model by \citet{2005A&A...436..895G} and estimated the number of cluster stars in each region as $N_{\rm i}=N_{\rm i}^{\rm obs}-N_{\rm i}^{\rm field}$, where $N_{\rm i}^{\rm obs}$ is the number of stars observed in each region and corrected for completess, and $N_{\rm i}^{\rm field}$ is the number of field stars predicted by the Galactic model in the same direction of NGC\,2808 and in a field of view with the same area as the outer field.

Since the number of binaries strongly depends on the number of stars in each sequence, and vice versa, in order to derive the unknowns of Eq.~\ref{eq:1} we applied an iterative procedure. 
 We started by assuming a null binary fraction and determine a crude estimate of $N_{\rm bMS}$, $N_{\rm mMS}$, and $N_{\rm rMS}$ by solving the system of Eq.~\ref{eq:1} for i=1,2,3.

 These numbers have been then used to determine $f^{\rm BIN}_{\rm i}$. To do this we have generated a CMD made of pure binaries as in \citetalias{2012A&A...537A..77M} by using a flat mass-ratio distribution and assuming that the binary components belong to any of the MSs.  The fraction of binaries, $f^{\rm bMS (mMS, rMS)}_{\rm i}$,  in each region $R_{\rm i}$, has been determined by computing the ratio between the total number of inserted binaries and the number of binaries in each region. We refer the reader to \citetalias{2012A&A...537A..77M} (see their Sect.~5.2 and 5.3) for details.

At this stage, we have obtained a raw estimate of $f^{\rm BIN}_{\rm i}$ and calculated $f^{\rm bMS}_{\rm i}$, $f^{\rm mMS}_{\rm i}$, and $f^{\rm rMS}_{\rm i}$. So we can derive $f^{\rm BIN}$ from Eq.~\ref{eq:1}. This ends one iteration. 

The values of  $N_{\rm bMS}$, $N_{\rm mMS}$, $N_{\rm rMS}$ and $f^{\rm BIN}$ are used to simulate a new CMD, improve the estimate of $f^{\rm BIN}_{\rm i}$, and again solve the system of Eq.~\ref{eq:1} for $N_{\rm bMS}$, $N_{\rm mMS}$, $N_{\rm rMS}$, and  $f^{\rm BIN}$. Following \citetalias{2012A&A...537A..77M}, we iteratively repeated the procedure until the value $f^{\rm BIN}$ changes by less than 0.001 from one iteration to the successive one.

Each measure is affected by the uncertainties in the determination of the fiducial line and on the corresponding boundaries of the CMD regions $R_{1}-R_{4}$. In order to determine it, we have repeated the procedure described above 1,000 times  each time by using different fiducials and the boundaries of the four regions, thus determining 1,000 values for the population ratio. Each fiducial has been determined by adding to each point of the fiducial a shift in color, whose value is randomly extracted by a Gaussian distribution with a $\sigma$ equal to the observed error. We assumed as the uncertainties in the determination of the fiducial line the $68^{\rm th}$ percentile of the 1,000 determination of the population ratio.
This uncertainty has been then added in quadrature to the Poisson uncertainty to determine the error associated to each measure.

In the central field, the three MSs are not distinguishable below $m_{\rm F814W}=21.25$ so we limited the analysis to the F814W magnitude range [$19.5-21.25$].  
In both the outer and middle field, where deep F475W and F814W photometry is available, we have analyzed the interval with $19.5<m_{\rm F814W}<22.0$ in close analogy with \citetalias{2012A&A...537A..77M}. In order to compare results from different fields in the outer and middle field we also provide results for the interval [$19.5-21.25$].

\begin{table*}\footnotesize
  \centering
\scriptsize {
    \caption {Fraction of stars in the three main stellar populations of NGC\,2808 derived from multiple MSs and RGBs in the three fields analyzed in this paper. The magnitude interval is also listed. The fraction of rMS, mMS, and bMS stars determined as described in Sect.~\ref{sec:ms} and the fraction of stars of RGB-(A+B+C), RGB-D, and RGB-E from \citetalias{2015ApJ...808...51M} are listed in the six upper lines of the table. The corresponding fraction of RGB stars inferred from method I are provided at line seven. The last four lines of the table list the population ratios derived from method II. See text for details. }
    \begin{tabular}{@{}ccccccc@{}} 
    \hline
field & sequence & $m_{\rm F814W}$ interval &  population ratio  &  population ratio  &  population ratio & binary fraction \\ 
      &          &                        & rMS or RGB-(A+B+C) & mMS or RGB-D & bMS or RGB-E & \\
\hline
outer    & MS  & 19.50-21.25 &   0.63$\pm$0.04 & 0.27$\pm$0.04 & 0.10$\pm$0.03 & 0.03$\pm$0.01 \\
         &     & 19.50-22.00 &   0.62$\pm$0.04 & 0.27$\pm$0.04 & 0.11$\pm$0.03 & 0.03$\pm$0.01 \\
middle   & MS  & 19.50-21.25 &   0.62$\pm$0.02 & 0.25$\pm$0.02 & 0.13$\pm$0.03 & 0.05$\pm$0.01 \\
         &     & 19.50-22.00 &   0.62$\pm$0.02 & 0.24$\pm$0.02 & 0.14$\pm$0.03 & 0.05$\pm$0.01 \\
inner, R$>$0.75 arcmin & MS  & 19.50-21.25 &   0.53$\pm$0.02 & 0.29$\pm$0.02 & 0.18$\pm$0.02 & 0.06$\pm$0.01\\
inner & RGB & 12.40-17.20 &   0.50$\pm$0.03 & 0.31$\pm$0.02 & 0.19$\pm$0.02 & -- \\
\hline
inner & RGB & 17.20-12.40 &   0.52$\pm$0.03  &  0.28$\pm$0.02  &  0.20$\pm$0.02 & --\\
\hline
outer                  & MS  & 19.50-22.00 &   0.57$\pm$0.04 & 0.30$\pm$0.04 &  0.13$\pm$0.03  & --\\
middle                    & MS  & 19.50-22.00 &   0.57$\pm$0.02 & 0.27$\pm$0.02 &  0.16$\pm$0.03 & --\\
inner, R$>$0.75 arcmin & MS  & 19.50-21.25 &   0.47$\pm$0.02 & 0.32$\pm$0.02 &  0.21$\pm$0.02 & -- \\
inner                  & RGB & 12.40-17.20 &   0.46$\pm$0.02 & 0.33$\pm$0.01 &  0.21$\pm$0.01 & -- \\
    \hline
   \end{tabular}
   \label{tab:results}
   }
\end{table*}
 
\begin{table}
  \centering
\scriptsize {
    \caption {Stellar masses for the three main populations of NGC\,2808 at different luminosities in the F814W band as inferred from BaSTI isochrones.}
    \begin{tabular}{cccc}
    \hline
    ${\it m}_{\rm F814W}\, {\rm [mag]}$ & ${\it M}_{\rm rMS}\, {\rm [M_{\odot}]}$ & ${\it M}_{\rm mMS}\, {\rm [M_{\odot}]}$ & ${\it M}_{\rm bMS}\, {\rm [M_{\odot}]}$ \\
    \hline
    $12.40$ & $0.856$ & $0.739$ & $0.654$ \\ 
    $17.70$ & $0.847$ & $0.732$ & $0.648$ \\ 
    $19.50$ & $0.769$ & $0.679$ & $0.610$ \\ 
    $21.25$ & $0.600$ & $0.537$ & $0.491$ \\ 
    $22.00$ & $0.529$ & $0.476$ & $0.435$ \\ 
    \hline
   \end{tabular}
   \label{tab:logmass}
   }
\end{table}

\section{Results}
\label{sec:results}

The obtained fractions of bMS, mMS and rMS stars and the fraction of binaries in each field are listed in Tab.~\ref{tab:results}. In the first two rows we have listed results corresponding to the F814W magnitude bin [$19.5-21.25$], while in Fig.~\ref{fig:risu} we plotted the fraction of b(m,r)MS-stars as a function of the average radial distance from the cluster center of all the analyzed stars.
  
Unfortunately, as discussed in Sect.~\ref{sec:data}, due to stellar crowding, the triple MS is not clearly visibile in the very central regions of NGC\,2808 and we have no information on the population ratio in the innermost 0.75 arcmin.
 In order to extend the study of the radial distribution of multiple stellar populations to the central regions, we have exploited the results from \citetalias{2015ApJ...808...51M}.

In that work, we have analyzed multi-wavelength photometry of NGC\,2808 as part of the UV Legacy Survey of Galactic Globular Clusters from GO-13297 and GO-10775. We have separated at least five distinct populations that we name A, B, C, D and E. The five populations are clearly visible along the RGB in the entire analyzed field of view, which corresponds to the inner field analyzed in this paper. 
Specifically, populations D and E correspond to the mMS and the bMS identified by \citet{2007ApJ...661L..53P}, while populations A, B, and C are associated to the rMS. 
 The fraction of stars in the five RGBs are $f^{\rm RGB-A}$=5.8$\pm$0.5\%, $f^{\rm RGB-B}$=17.4$\pm$0.9\%, $f^{\rm RGB-C}$26.4$\pm$1.2\%, $f^{\rm RGB-D}$=31.3$\pm$1.3\%, and $f^{\rm RGB-E}$=19.1$\pm$1.0\% of the total number of RGB stars with $12.25<m_{\rm F814W}<17.70$, respectively. Therefore the progeny of the rMS, which corresponds to the RGBs A, B, and C include $f^{\rm RGB-(A+B+C)}$=49.6$\pm$1.6\% of the total number of MS stars.
 The interval of luminosity analyzed in \citetalias{2015ApJ...808...51M} for the study of multiple RGBs obviously differs from that of this paper. In order to properly compare RGB and MS stars we have adopted two different methods.

\subsection{Crude estimate (Method I)}\label{sub:metI}
The first method is based on the comparison of results from multiple MS corresponding to the F814W magnitude bin [$19.5-21.25$] and from the RGB.
 Specifically, we have first derived the population ratios from four distinct sample of stars that include MS stars of the inner, middle, and outer field and RGB stars of the inner field.

To compare results from the RGB and the MS, we used the photometric catalog from \citetalias{2015ApJ...808...51M} and estimated the fraction of RGB-E, RGB-D and RGB-(A$+$B$+$C) as done in that paper but for stars with radial distance from the cluster center larger than 0.75 arcmin. This is the same region of the inner field where we have determined the fraction of bMS, mMS, and rMS stars. 

We found that the fraction of RGB-(A$+$B$+$C), RGB-D and RGB-E stars are $f^{\rm RGB-(A+B+C)}_{\rm inner-field, R>0.75'}=$0.51$\pm$0.03, $f^{\rm RGB-D}_{\rm inner-field, R>0.75'}=$0.32$\pm$0.02, $f^{\rm RGB-E}_{\rm inner-field, R>0.75'}=$0.17$\pm$0.02. 
For completeness, we derived the population ratio for stars with radial distance from the cluster center R$<$0.75 arcmin. We find $f^{\rm RGB-(A+B+C)}_{\rm inner-field, R<0.75'}=$0.49$\pm$0.03, $f^{\rm RGB-D}_{\rm inner-field, R<0.75'}=$0.30$\pm$0.02, $f^{\rm RGB-E}_{\rm inner-field, R>0.75'}=$0.21$\pm$0.03 and conclude that there is no evidence for significant difference in the population ratio derived within the region with radial distance $0'.00<$R$<0'.75$ and the region with $0'.75<$R$<2'.10$.
 
\begin{figure*}
\centering
   \includegraphics[width=5.75cm]{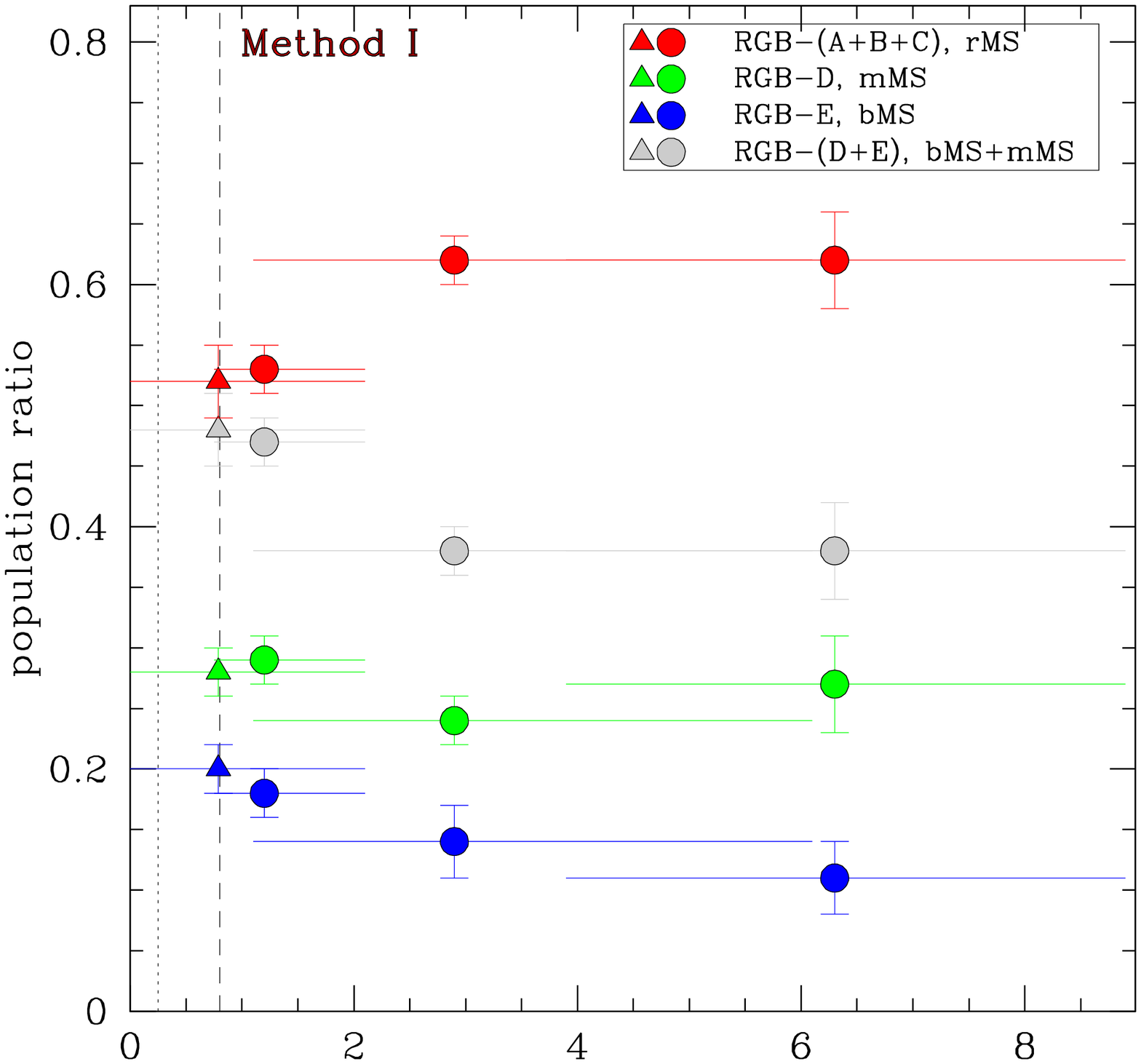}   
   \includegraphics[width=5.75cm]{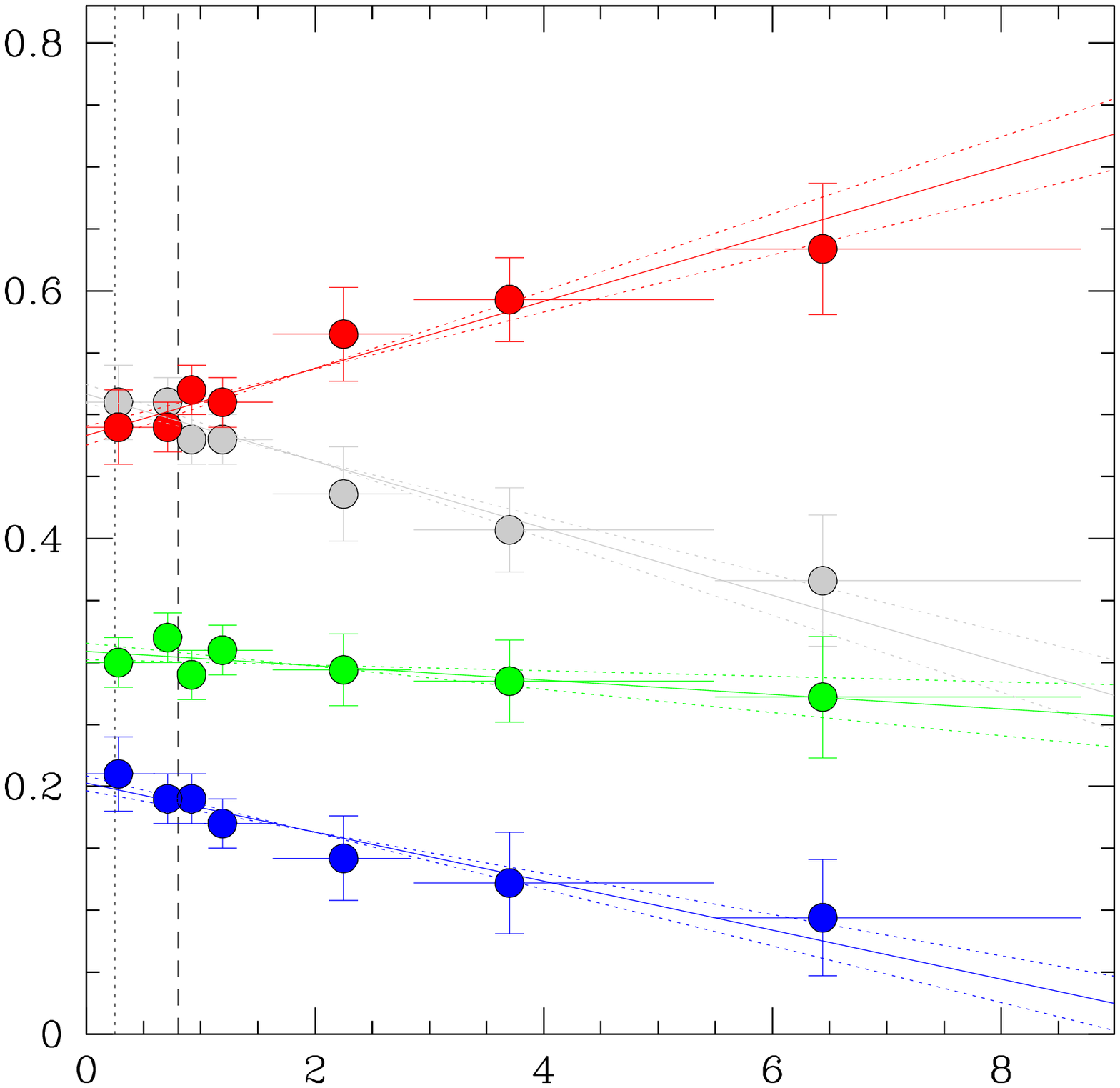}\\ 
   \includegraphics[width=5.75cm]{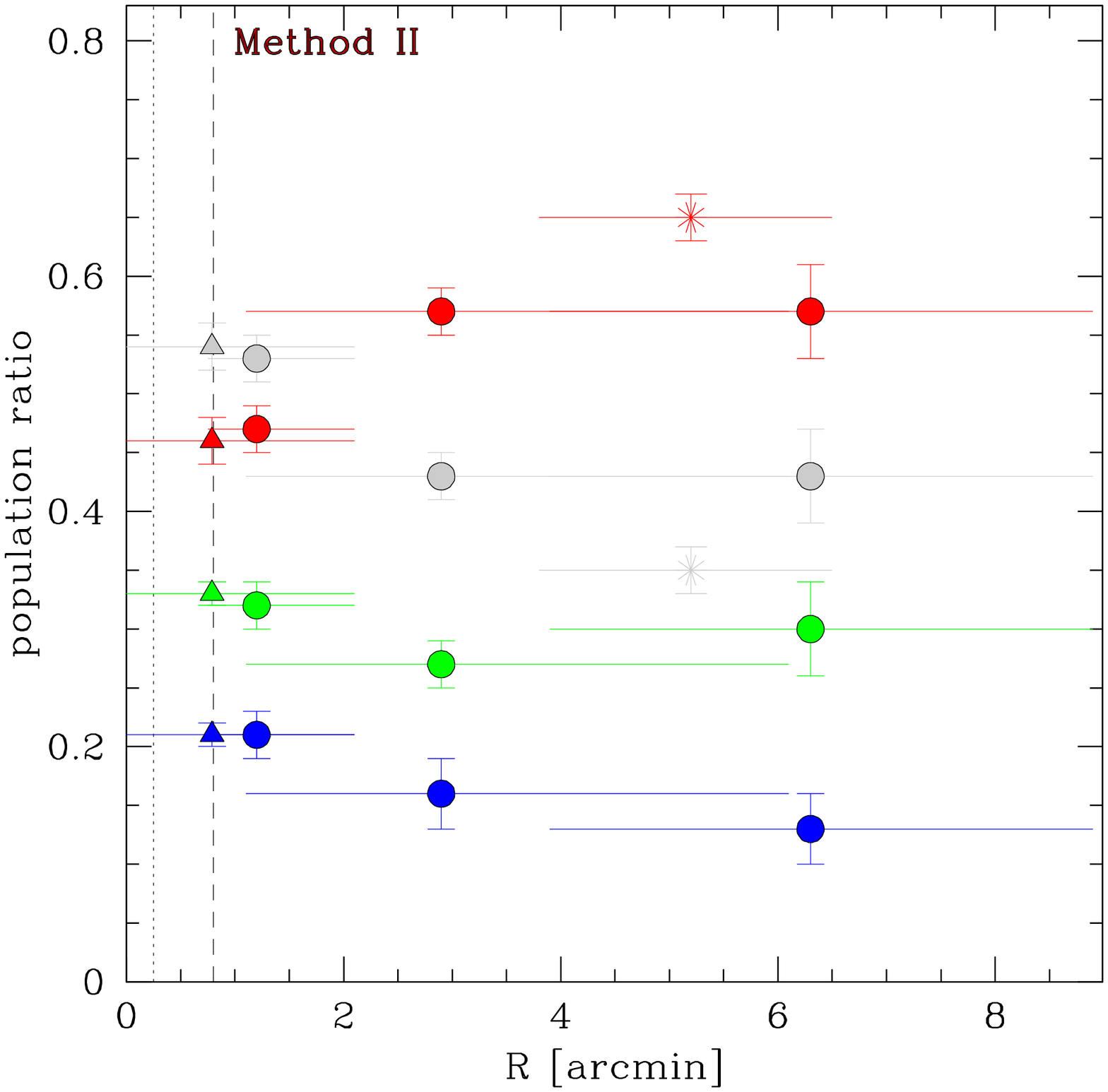}   
   \includegraphics[width=5.75cm]{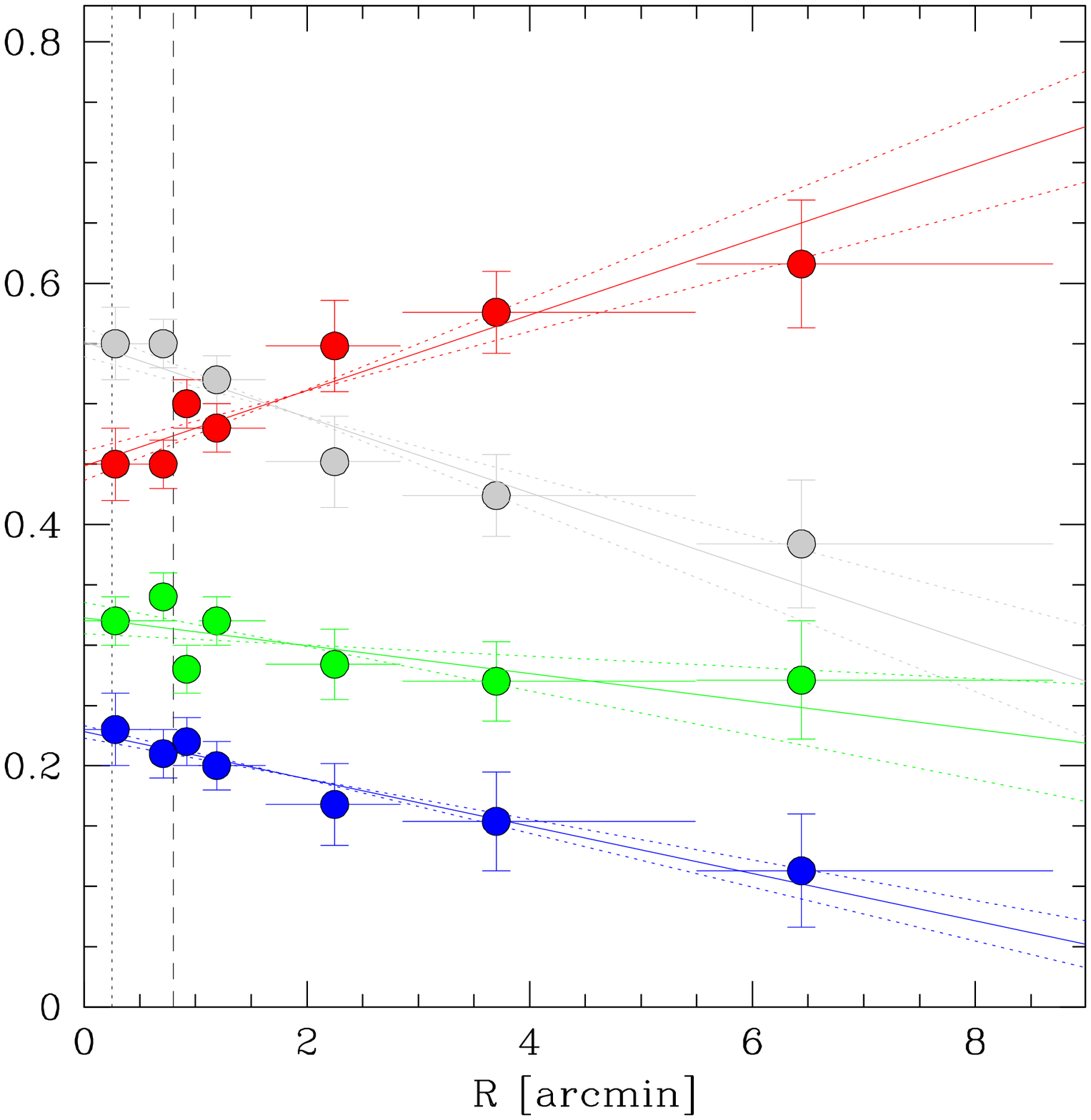}   
   \caption{Fraction of bMS (blue symbols), mMS (green symbols), rMS (red symbols) and bMS$+$mMS (gray symbols) stars respect to the total number of MS stars as a function of the radial distance from the cluster center. The triangles used in the left panels indicate the values of population ratios inferred from RGB stars, while the squares refer to results from the MS. The red and gray asterisks plotted in the lower-left panel indicate the population ratio inferred by \citet{2012ApJ...754L..34M} in their study of low-mass MS stars.  The horizontal lines plotted in each panel span the radial covarage of each bin, while the squares and the triangles indicate the average radial distance of the stars in each region. The vertical dotted and dashed lines mark the core and the half-light radius, respectively. 
Upper and lower panels show results obtained from method I and method II, respectively.
 In the upper-left and lower-left panel we have analyzed each field independently, while the upper-right and lower-right panel show the results for seven region with different distance from the cluster center. The best-fit straight lines are shown in the right panels, where dotted lines represent the maximum- and minimum-slope straight lines.
}
   \label{fig:risu}
\end{figure*}

 We thus imposed the same fraction of stars along the three MSs and the corresponding RGBs in the region of the inner field with R$>$0.75 arcmin and calculated the number of RGB-(A$+$B$+$C) stars as:\\
$N^{\rm RGB-(A+B+C)}_{\rm inner-field}$=$N^{\rm RGB-(A+B+C)}  \frac{f^{\rm rMS}_{\rm inner-field}} {f^{\rm RGB-(A+B+C)}_{\rm inner-field, R>0.75'}}$.\\
 We derived $N^{\rm RGB-D (E)}_{\rm inner-field}$ similarly. Results are listed in Tab.~\ref{tab:results} and illustrated in the upper-left panel of Fig.~\ref{fig:risu}.
 
The fraction of stars in each sequence in the middle and the outer field agree within $1\,\sigma$, where approximately 63\% of the entire number of the analyzed MS stars in the F814W magnitude interval [$19.5-21.25$] belong to the rMS. The fraction of rMS stars slightly decreases in the central field to about 53\%, where it is similar to the fraction of RGB-(A$+$B$+$C). The difference between the fraction of rMS stars in the inner and in the outer field is significant at the $2.5\,\sigma$-level and similar results are obtained when we compare results from the RGB in the inner field with those from the MS in the outer and middle fields. 

 In contrast, the fraction of mMS- plus bMS-stars seems to increase when moving from the cluster outskirts to the center, and both the blue and the middle MS are more centrally concentrated than the rMS.
 We note, that one of the main disadvantages of the analysis illustrated in the upper-left panel of Fig.~\ref{fig:risu} is that, in some cases, distinct fields cover the same radial interval and the large size of each bin has the effect of diluting and hiding in part any existing radial gradient.
 In order to further explore the radial distribution of the multiple stellar populations in NGC\,2808 and better identify the presence and strength of radial gradients, it is necessary to use a finer and non-overlapping binning.
 
In order to further investigate the radial distribution of multiple stellar populations in NGC\,2808 we have divided the stars studied in this paper into seven groups with different radial distance from the cluster center and calculated the population ratio in each region. 
Specifically, we have defined a circle with radius R$=0.60$ arcmin where photometry of RGB stars only is available and three additional regions included in the inner field such that, each region contains the same number of MS stars. Moreover, we have defined three additional annuli that include stars in the middle and the outer field. The boundaries of these seven regions are plotted with dotted circles in Fig.~\ref{fig:chart}. We have verified that the conclusion of the paper do not depend either on the location or on the number of regions that we use to study the radial distribution of multiple stellar populations. 

Results are listed in Table~\ref{tab:risu} and illustrated in the upper-right panel of Fig.~\ref{fig:risu}, where the continuous lines are the best-fit weighted-least-squares straight lines. The values of the slope $a$ and the intercept $b$ of each line are listed in Table~\ref{tab:slope1} together with the correspoding uncertainties.

The slope corresponding to the red MS is larger than zero with a significance greater than $6\,\sigma$, while both the blue and the middle exhibit negative values of $a$ with a significance of $\sim6.5\,\sigma$ and $\sim1.5\,\sigma$, respectively.
 By assuming a flat distribution, and the same uncertainties as in our population-ratio estimates, $10^6$ Monte Carlo simulations indicate we have a probability of $2\times 10^{-4}$ to get a slope equal to or higher than that observed for red MS stars.
  In the case of the bMS, the probability that the observed gradient is due to measurement errors is  $4\times 10^{-3}$.

Comparing red, middle and blue MS, it can be noted that the bMS appears to be the most concentrated, having the minimum slope value among them. The difference between the slopes associated to rMS and bMS results to be significant at $~\,12\sigma$. The mMS is also more concentrated than the rMS with a difference between the two slopes that is significant at $\sim\,7\sigma$.
  
\begin{table*}\footnotesize
\centering 
\scriptsize {
    \caption {Population ratios measured in the redefined radial sampling (see text for details); ${\rm R}_{\rm min}$, ${\rm R}_{\rm max}$, and ${\rm R}_{\rm med}$ are respectively the minimum,maximum and median radial distances of the bin. The first lines ar referred to Method I while the lower part is referred to Method II.}
    \begin{tabular}{ccccccc}
    \hline
    ${\rm R}_{\rm min}$ & ${\rm R}_{\rm max}$ & ${\rm R}_{\rm med}$ & Population ratio & Popution ratio & Population ratio & Population ratio\\
    $[{\rm arcmin}]$    & $[{\rm arcmin}]$    & $[{\rm arcmin}]$    & rMS              & mMS            & bMS              & mMS+bMS         \\
    \hline
    $0.00$ & $0.60$ & $0.28$ & $0.49\pm 0.03$   & $0.30\pm 0.02$   & $0.21\pm 0.03$   & $0.51\pm 0.036$\\ 
    $0.60$ & $0.82$ & $0.71$ & $0.49\pm 0.02$   & $0.32\pm 0.02$   & $0.19\pm 0.02$   & $0.51\pm 0.028$\\ 
    $0.82$ & $1.03$ & $0.92$ & $0.52\pm 0.02$   & $0.29\pm 0.02$   & $0.19\pm 0.02$   & $0.48\pm 0.028$\\ 
    $1.03$ & $1.63$ & $1.19$ & $0.51\pm 0.02$   & $0.31\pm 0.02$   & $0.17\pm 0.02$   & $0.48\pm 0.028$\\ 
    $1.63$ & $2.84$ & $2.25$ & $0.565\pm 0.038$ & $0.294\pm 0.029$ & $0.142\pm 0.034$ & $0.436\pm 0.045$\\ 
    $2.86$ & $5.49$ & $3.70$ & $0.593\pm 0.034$ & $0.285\pm 0.033$ & $0.122\pm 0.041$ & $0.407\pm 0.053$\\ 
    $5.50$ & $8.70$ & $6.44$ & $0.634\pm 0.053$ & $0.272\pm 0.049$ & $0.094\pm 0.047$ & $0.366\pm 0.068$\\ 
    \hline
    $0.00$ & $0.60$ & $0.28$  & $0.45\pm 0.03$   & $0.32\pm 0.02$   & $0.23\pm 0.03$   & $0.55 \pm 0.036$\\ 
    $0.60$ & $0.82$ & $0.71$  & $0.45\pm 0.02$   & $0.34\pm 0.02$   & $0.21\pm 0.02$   & $0.55 \pm 0.028$\\ 
    $0.82$ & $1.03$ & $0.92$  & $0.50\pm 0.02$   & $0.28\pm 0.02$   & $0.22\pm 0.02$   & $0.50 \pm 0.028$\\ 
    $1.03$ & $1.63$ & $1.19$  & $0.48\pm 0.02$   & $0.32\pm 0.02$   & $0.20\pm 0.02$   & $0.52 \pm 0.028$\\ 
    $1.63$ & $2.84$ & $2.25$  & $0.548\pm 0.038$ & $0.284\pm 0.029$ & $0.168\pm 0.034$ & $0.452 \pm 0.045$\\ 
    $2.86$ & $5.49$ & $3.70$  & $0.576\pm 0.034$ & $0.270\pm 0.033$ & $0.154\pm 0.041$ & $0.424 \pm 0.053$\\ 
    $5.50$ & $8.70$ & $6.44$  & $0.616\pm 0.053$ & $0.271\pm 0.049$ & $0.113\pm 0.047$ & $0.384 \pm 0.068$\\ 
    \hline
   \end{tabular}
   \label{tab:risu}
   }
\end{table*}

\begin{table}
\centering 
\scriptsize {
    \caption {Slope and intercept values ($a$ and $b$ respectively), together with their uncertainties ($\sigma_{\rm a}$ and $\sigma_{\rm b}$), of best-fit least-squares lines for rMS, mMS, bMS and mMS+bMS respectively, referrend to Method I; see text for details.}
    \begin{tabular}{ccccc}
    \hline
    sequence & $a$ & $\sigma_{\rm a}$ & $b$ & $\sigma_{\rm b}$ \\ 
    \hline
    rMS     & $ 0.027$ & $0.004$ & $0.483$ & $0.007$\\ 
    mMS     & $-0.006$ & $0.003$ & $0.309$ & $0.007$\\ 
    bMS     & $-0.020$ & $0.003$ & $0.202$ & $0.006$\\ 
    mMS+bMS & $-0.027$ & $0.004$ & $0.517$ & $0.007$\\ 
    \hline
    \end{tabular}
    \label{tab:slope1}
   }
\end{table}

\subsection{A more sophisticated estimate (Method II)}\label{sub:metII}
As a second method to compare results from the RGB and from the MS we follow the receipe by Milone et al.\,(2009) in their study of the radial distribution of stellar populations along the SGB of NGC\,1851.
Sure enough, stars in different intervals of luminosity (like the sample of RGB and MS stars analyzed in this paper) have different masses; moreover, stars with different luminosity but different helium abundance have different masses.
This second method takes in account this mass difference in order to properly compare the population ratio inferred from the MS and the RGBs.
We have to compensate for the  fact that the two stellar groups  that define the distinct MSs and RGBs span different  mass intervals.

 This method allows us to fully exploit information from the entire magnitude interval with $19.5<m_{\rm F814W}<22.0$ in the middle and outer field. We have normalized the number of stars in each sequence to the analyzed mass interval as: \\
$N_{\rm b(m,r)MS}/\Delta\mathcal{M}$
where
$\Delta\mathcal{M}=\int_{M1,\rm b(m,r)MS}^{M2, \rm b(m,r)MS} \phi(\mathcal{M})d\mathcal{M}$.\\
Here, $\phi(\mathcal{M})$ is the adopted mass function, and $M1,\rm b(m,r)MS$ and $M2, \rm b(m,r)MS$ are the minimum and the maximum stellar mass in the analyzed interval of luminosity. 
We adopted a similar relation to derive the the normalized fraction of stars in the RGB-(A$+$B$+$C), RGB-D, and RGB-E.

 The masses corresponding to different luminosities are listed in Tab.~\ref{tab:logmass} and are obtained from  BaSTI isochrones \citep{2004ApJ...612..168P,2009ApJ...697..275P} by adopting the same values of distance modulus, reddening, age, metallicity and helium abundance as in \citetalias{2012A&A...537A..77M}.
 We adopted for the mass function $\phi(\mathcal{M})$=$M^{\alpha}$, where we used the values of $\alpha$ derived by \citetalias{2012A&A...537A..77M}. 
 Results are listed in Tab.~\ref{tab:results} and shown in the lower panels of Fig.~\ref{fig:risu}.

In the lower-left panel we have considered each field separately.
We confirm results obtained by using the method I, with the population corresponding to the rMS, being less centrally concentrated than the stellar populations of NGC\,2808 that are highly helium enhanced. In this case the difference between the fraction of RGB-(A+B+C) stars derived in the central field ($\sim$57\%) and the fraction of rMS in the middle and the outer fields ($\sim$46-47\%) is significant at the $\sim4.0\,\sigma$ and $\sim2.5\,\sigma$ level, respectively.
 
 A recent study, based on images taken with the near-infrared (NIR) channel of WFC3 for stars in three fields have investigated multiple sequences of very low mass stars in NGC\,2808. The three analyzed NIR/WFC3 fields have all radial distance of $\sim$5.2 arcmin from the center of NGC\,2808 (\citealt{2012ApJ...754L..34M}; red dotted fields in Fig.~\ref{fig:chart}).  
 Two MSs are clearly visible in the magnitude interval with $<21.25<m_{\rm F160W}<22.50$. The most-populous ones, contains 65$\pm$2\% of the total number of analyzed stars and corresponds to the red MS, while the remaining 35$\pm$2\% the sequence associated to the blue and the middle MS. The population ratios derived by Milone et al.\,(2012) and normalized to the corresponding mass intervals are in agreement with the results from this paper and are represented with red and grey asterisks in the lower-left panel of Fig.~\ref{fig:risu}. 
 
In the lower-right panel of Fig.~\ref{fig:risu} we show the results for the seven regions with different radial coverage, in close analogy with what we have done in the upper-right panel. The slopes corresponding to the three sequences are listed in Table~\ref{tab:slope2} and confirm the conclusion of Sect.~\ref{sub:metI}.
In this case Monte Carlo simulations provide a probability smaller than 1$\times 10^{-4}$ to get a slope equal to or higher than that observed for red MS stars, while the probability that the observed gradient for bMS stars is due to measurement errors is  3$\times 10^{-3}$.
The difference between mMS and rMS slopes is now significant at $\sim\, 5 \sigma$ while the significance level of the difference between bMS and rMS slopes is $\sim\, 7\sigma$. 

\begin{table}
\centering 
\scriptsize {
    \caption {Slope and intercept values ($a$ and $b$ respectively), together with their uncertainties ($\sigma_{\rm a}$ and $\sigma_{\rm b}$), of best-fit least-squares lines for rMS, mMS, bMS and mMS+bMS respectively, referrend to Method II; see text for details.}
    \begin{tabular}{ccccc}
    \hline
    sequence & $a$ & $\sigma_{\rm a}$ & $b$ & $\sigma_{\rm b}$ \\ 
    \hline
    rMS     & $ 0.031$ & $0.006$ & $0.449$ & $0.012$\\ 
    mMS     & $-0.011$ & $0.007$ & $0.322$ & $0.013$\\ 
    bMS     & $-0.020$ & $0.003$ & $0.228$ & $0.005$\\ 
    mMS+bMS & $-0.031$ & $0.006$ & $0.551$ & $0.012$\\ 
    \hline
   \end{tabular}
   \label{tab:slope2}
   }
\end{table}

\subsection{Theoretical interpretation}\label{sub:theor}
The results of our observational analysis show that two helium enhanced populations (bMS and mMS) are more centrally concentrated than the rMS population.
This result is in general qualitative agreement with the predictions of multiple-population cluster formation models according to which second-generation (2G) stars should form more concentrated than the initial first-generation (1G) population. 

Because of the effects of dynamical evolution on the structural properties of the various stellar populations, a direct connection between the current observed properties and those predicted by cluster formation models is not straightforward. 
The long-term cluster dynamical evolution will gradually weaken the initial radial gradient in the fraction of 2G stars (see e.g.\,\citealt{2013MNRAS.429.1913V}) until complete spatial mixing when no memory of the initial differences is preserved and all the populations share the same spatial distribution.

As discussed in the Intoduction, a few observational studies have found clusters still retaining memory of the initial differences in the spatial distribution of 1G and 2G populations while others appear to have reached the phase when different populations are completely mixed. 
For those clusters for which a radial gradient in the fraction of 2G stars is still present, the strength of the current observed radial gradient provides a lower limit on what must have been a stronger initial gradient. 

A complete and detailed model for any specific individual cluster is a very challenging and computationally expensive task (see e.g.\,\citealt{2011MNRAS.410.2698G}, \citealt{2014MNRAS.445.3435H}); the presence of multiple populations with the additional complexities related to their formation and dynamical history further complicates this task. This is well beyond the scope of the goals of this paper and is deferred to future investigations.
Here, in order to illustrate the process of spatial mixing of multiple stellar populations and provide some initial insight on the possible dynamical history leading to the gradient found in our analysis we present the results of a simple N-body simulation focussing our attention on the two-body relaxation-driven long-term evolution of the spatial distributions of the 1G and 2G population.
We started our simulation with 50000 particles with the 1G and the 2G populations each having half of the total mass of the system and a range of stellar masses equal to those expected at about 12 Gyr for a system with a \citet{2001MNRAS.322..231K} stellar IMF. Both populations are characterized by a \citet{1966AJ.....71...64K} density profile with central dimensionless potential $W_{0}=7$ but the 2G population is initially concentrated in the inner regions of the cluster and has a half-mass radius about 4.5 times smaller than that of the 1G population. 
The simulation was run on the Big Red II supercomputer at Indiana University with the GPU-accelerated version of the NBODY6 code (\citealt{2003gnbs.book.....A}, \citealt{2012MNRAS.424..545N}).

The cluster is initially tidally limited and assumed to be on a circular orbit in the external potential of a host galaxy modeled as a point mass. We focus here solely on the long-term evolution driven by two-body relaxation and its effect on the spatial distributions of the two populations (see also \citealt{2013MNRAS.429.1913V} for further discussion on the long-term evolution and dynamics of spatial mixing). 

\begin{figure}
\centering
   \includegraphics[width=9cm]{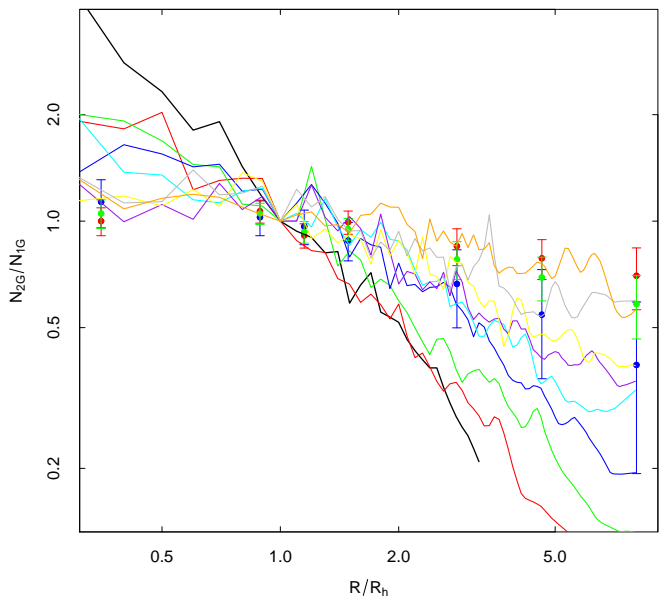}
   \caption{Time evolution of the radial profile of the ratio of the number of 2G to 1G stars (normalized to the value measured at the half-mass radius) versus the projetced distance from the cluster center (normalized to the half-mass radius) for the simulation discussed in the paper (see Section~\ref{sub:theor}). Different curves correspond to the profiles measured at values of $t/t_{\rm rh}(t)$ equal to about $0.0$ (black line), $2.0$ (red line), $3.0$ (green line), $5.0$ (blue line), $7.0$ (cyan line), $8.5$ (purple line), $10.3$ (yellow line),$12.5$ (grey line), $15.3$ (orange line). Each radial profile has been calculated by using nine snapshots around the indicated time. Points with error bars show the observed ratio of the number of bMS to rMS stars (blue points), mMS stars to rMS stars (red points) and (bMS+mMS) to rMS stars (green points).
}
   \label{fig:simu}
\end{figure}

In Fig.~\ref{fig:simu} we show the time evolution of the radial profile of the ratio of the number of 2G to 1G stars for stars with masses between 0.6 and 0.8 ${\rm M_{\odot}}$ along with the observed ratios for $N_{\rm bMS}/N_{\rm rMS}$, $N_{\rm mMS}/N_{\rm rMS}$ and $N_{\rm bMS+mMS}/N_{\rm rMS}$. The results of the simulations illustrate the progressive weakening of the initial radial gradient. In particular, the radial gradient in the fraction of bMS and mMS stars currently observed in NGC\,2808 is approximately consistent with those found in the simulation at times $t/t_{rh}(t)\geq 7$. This figures shows a possible dynamical path and the significantly stronger initial gradient behind the current structural properties of the multiple populations of NGC\,2808.

As already discussed in the Introduction, NGC\,2808 is a particularly complex cluster; the early dynamics during the sequence of events that led to the formation of the different populations observed in this cluster are still unclear. A more detailed exploration of the possible differences in the early and long-term dynamical evolution of different populations and the implications for the current observed properties will require a much more extensive study than that presented here.

\section{Summary and Conlusions}
NGC\,2808 is one of the most-massive GCs in the Galaxy and hosts at least five distinct stellar populations, namely A, B, C, D, and E, with different content of helium and light elements (\citetalias{2015ApJ...808...51M}). 
 Populations D and E are highly enhanced in helium up to $Y \sim$0.32 and $\sim$0.38 and correspond to the mMS and the bMS discovered by \citet{2007ApJ...661L..53P}. Populations A, B, and C are connected with the red MS by \citet{2007ApJ...661L..53P} and have almost primordial helium abundance.
 In this paper we have used both archive and proprietary data collected with the ACS/WFC and WFC3/UVIS on board of {\it HST} to investigate the radial distribution of the three main populations of NGC\,2808 which correspond to the bMS, the mMS, and the rMS. Our dataset includes three fields spanning a radial interval that ranges from the cluster center to approximately 8.5 arcmin.
 
 Parallel ACS@HST observations taken as part of GO-12605 are presented here for the first time (see upper-left panel of Fig.~\ref{fig:CMDs} for the obtained CMD).
The three MSs have been detected in all the analyzed fields. The fraction of stars in each main sequence has been determined starting from a radial distance of 0.75 arcmin from the cluster center to 8.5 arcmin. At radial distance smaller than 0.75 arcmin, the three MS can not be clearly distinguished due to stellar crowding.  
 We have used the photometry of RGB stars from \citetalias{2015ApJ...808...51M} to extend the study of multiple stellar populations to the cluster center.
 Using two different methods, we found that the populations which correspond to the rMS are less centrally concentrated than the helium rich stellar populations, with a significance for the result that is higher than $3\sigma$.
We have also presented the results of a simple N-body simulation illustrating the possible evolution of the multiple population spatial mixing of this cluster.\\

NGC\,2808 has a very extended HB which is well populated on both sides of the RR\,Lyrae instability strip \citep{1997ApJ...480L..35S}. The red HB of NGC\,2808 shares the same chemical composition as the stellar populations corresponding to the rMS stars (\citealt{2013A&A...549A..41G}; \citealt{2014MNRAS.437.1609M}). The blue MS corresponds to the bluest HB tail, while the remaining blue-HB stars are connected with the middle MS (e.g.\,\citealt{2005ApJ...631..868D}; \cite{2007ApJ...661L..53P}; \citealt{2011MNRAS.410..694D}).
 \citet{2000A&A...363..159B} have investigated the radial distribution of the HB compontents and find no evidence for a significant gradient.
 \citet{2009ApJ...696L.120I} further analyzed the radial distribution of HB stars in NGC\,2808 and suggested that red-HB stars are less centrally concentrated than the remaining HB stars; although their conclusion is significant only at $1.5\sigma$ level, it suggests the presence of a gradient consistent with that we find in our analysis.

Recent studies have investigated the properties of the triple MS in NGC\,2808, like the luminosity and mass function and the internal kinematics. 
\citetalias{2012A&A...537A..77M} have studied the mass functions of the three MSs discovered by \citet{2007ApJ...661L..53P} and found that the slope of rMS-, mMS-, and bMS-mass function are  $\alpha=-1.2\pm0.3$, $\alpha = -0.9 \pm 0.3$, and $\alpha = -0.9 \pm 0.4$, rispectively, i.e.\,are the same the same within the errors.
In a paper from this series, \citet{2015ApJ...810L..13B}, have investigated the internal kinematics of the stellar populations in NGC\,2808 by using the same dataset from the central field used in this paper. They have found that in the most-external region that they have analyzed, between $1.5$ and $\sim$2.0 times the half-light radius, the proper-motion distributions of the populations  D and E, are significantly more anysotropic than that of the populations A, B, and C. On the basis of results from N-body simulation, Bellini and collaborators have suggested that the kinematic difference between the populations highly enhanced in helium and those with with almost primordial helium, are consistent with a scenario, where populations D and E were more-centrally concentrated at the time of their formation.

\label{sec:summary}

\section*{Acknowledgements}
We thank the anonymous referee for her/his comments that improved the quality of the present work.
M.S., A.A. and G.P. acknowledge support from the Spanish Ministry of Economy and Competitiveness (MINECO) under grant AYA2013-42781.
M.S. and A.A. acknowledge support from the Instituto de Astrof\'{i}sica de Canarias (IAC) under grant 309403.
G.P. acknowledge partial support  by the Universit\`a degli Studi di Padova Progetto di Ateneo CPDA141214 ``Towards understanding complex star formation in Galactic globular clusters'' and by INAF under the program PRIN-INAF2014.
E.V. and J.H. acknowledge support from STScI grant GO-13297.

\bibliographystyle{aa}

\label{lastpage}
\end{document}